
\baselineskip=16.2pt\magnification=1200
\def\n{\noindent}
\def\s{\sin\theta}
\def\ss{\sin^2\theta}
\def\c{\cos\theta}
\def\r{$\rightarrow$}
\def\i{\item}
\line{\hfill IUCAA-3/95}

\vskip.3truein

\centerline{\bf EVOLUTION OF INTRA-CAVITY FIELDS AT NON-STEADY STATE}
\centerline{\bf IN DUAL RECYCLED INTERFEROMETER}
\vskip.6truein
\centerline{\bf Biplab Bhawal}
\centerline{Inter University Centre for Astronomy and Astrophysics,}
\centerline{Post Bag 4, Ganeshkhind,}
\centerline{Pune- 411007, INDIA}
\vskip.2truein
\centerline{(Email: biplab@iucaa.ernet.in)}
\vskip.5truein
\centerline{\bf Abstract}
{\sl We describe how exactly the intra-cavity fields in a dual recycling
cavity build up their power before achieving a steady state value. We
restricted our analysis here to interferometers with lossless mirrors
and beam-splitter. The  complete series representation of the intra-cavity
lights at any stage of evolution in non-steady state have been presented.}
\vskip1.0truein

\centerline{\bf To be published in Applied Optics}

\vfill\eject

Laser interferometric gravitational wave observatories[1] are currently
being designed or constructed in several countries (LIGO, VIRGO, GEO,
AIGO). The interferometer, in essence, is a rather elaborate transducer
from optical path difference to output power. So, monitoring the change
in output power, it would be possible to detect the changing curvature
of spacetime induced by the passage of a gravitational wave. In order to
reduce the shot noise level, an integral feature of these detectors will be
the use of high power laser, in conjunction with variants of optical
technique known as light recycling. The first one of these techniques is
called {\it power recycling}[2], in which, at a dark fringe operation of
the interferometer, the outgoing laser light is recycled back into the
interferometer by putting a mirror in front of the source, thus enhancing
the laser power. When a gravitational wave passes through an interferometer,
it modulates the phase of the laser light, thus producing sidebands which
travel towards the photodetector [3]. So, another variant of optical
technique called {\it dual recycling} incorporates a second recycling
mirror placed in front of the photodetector as shown in Fig.1. This
arrangement can store signal sidebands for sufficiently long time to
allow optimum photon noise sensitivity within a restricted bandwidth [4].

In this note, we describe, how with the help of {\it Mathematica}[5], one
can do  calculation for the evolution of intra-cavity fields in the two
recycling cavities of a dual recycled interferometer and finally achieve
the expressions for the intra-cavity fields at the steady state condition.
This may find its application in investigating various kinds of transient
states in such interferometers.

We first introduce the notations used here. In all the following figures,
rays of light labelled by $a$, $b$, $x$, $y$
represent the complex amplitude of the beam's electromagnetic field.

\n{\bf Mirrors:} In this note, we make the following assumptions on the
mirrors and beam-splitter: (i) these are lossless, (ii) the substrate and
the dielectric layers in mirrors are linear media, (iii)
mirrors are time-reversal invariant, (iv) any mirror has a reflection
symmetry with respect to exchange of inputs and outputs on the same side
of the mirror, (v) there is no differential time delay between the
transmitted and reflected lights and also the overall time delay across
the mirror is negligible. One can then arrive at the following simple
input-output mirror relation (refer to Fig.2):
$${a_1 \choose b_1}=\left(\matrix{t&-r\cr r&t}\right){b_0\choose a_0},
\eqno(1)$$
where $t$ and $r$ represent the transmission and reflection coefficients
of the mirror. For a lossless mirror, $\mid t\mid^2+\mid r\mid^2=1$.
We consider all the beam-splitters to be of 50:50 type.

\n{\bf Fabry-Perot Cavity:}  The flowchart of the light paths in such a
cavity has been shown in Fig.3. Light beams $a_0$ and $b_0$ enter the
cavity from corner mirror (CM) and end mirror (EM) sides respectively.
The beam $x$ represents the intra-cavity field that grows up in every
bounce it gets from CM or EM, provided it adds up in phase with the
incoming light there. The index $j$ represents the number of round-trips
of the $x$-field inside the cavity. The circle with the quantity $L$
inside it represents the change in phase of the intra-cavity light due
to the traversal of  cavity of length $L$. After a sufficient number of
bounces, the intra-cavity field, $x$, achieves a steady state when it
stops growing any more (i.e., the change after any bounce would be
negligibly small).

\n{\bf Dual Recycling Cavity:} The corresponding flowchart for a single
delay-line dual-recycled cavity has been shown in Fig.4. To understand
this figure one may also refer to Fig.1.
The circle with the quantity $l_1$ or $l_2$ inside it represents the
change in phase of the intra-cavity light due to traversal of power
recycling cavity (PRC) of length $l_1$  between the beam-splitter (BS)
and the power recycling mirror (PRM) or Signal Recycling Cavity (SRC) of
length $l_2$ between BS and signal recycling mirror (SRM) respectively.
The electromagnetic field $a_0$ represents the coherent beam of laser light
 entering through PRM. The field $b_0$ is in vacuum state in ordinary
 interferometers, but as suggested by Caves[6], in advanced interferometers,
this may be made to be in squeezed vacuum state to reduce the shot noise.
The  fields $x_4$ and $y_4$ represent intra-cavity light beams in the
Power Recycling Cavity (PRC) and Signal Recycling Cavity (SRC) respectively.
We represent the amplitude transmission and reflection coefficients of the
PRM (SRM) by $t_1$($t_2$) and $r_1$($r_2$) respectively. The box with the
label `Propagation in Arms' represents the matrix relation:

$${x_3[j]\choose y_3[j]}=\exp\bigg[{i2\omega_0 L_a\over c}\bigg]
\left(\matrix{e^{-i\theta}&0\cr 0&e^{+i\theta}\cr}\right)
{x_2[j]\choose y_2[j]},\eqno(2)$$
where $k_0=\omega_0/c$ and $\omega_0$, $L_a$, $c$ are the circular frequency
of light beam, the arm length of the interferometer and the velocity of
light respectively. The quantity $\theta$ represents the differential
phase shift of light in the two arms of the interferometer. The end
mirrors of dual recycling cavity are assumed to be perfectly reflecting
ones.

 A small programme in {\it Mathematica} can be easily written to implement
this recursive procedure. After a sufficient number of bounces, $j$
(say, 10), one can look at the expressions for $x_4[j]$ and $y_4[j]$. We
have  set $x_0[0]=y_0[0]=0$ in the final expressions of $x_4[j]$ and
$y_4[j]$  after finishing the recursive calculation because the initial
intra-cavity fields were zero. Now, grouping the related terms together,
we can guess the formation of a series in each case.

The series obtained by us are as follows:
$$\eqalignno{
x_4[j]=& b_0 \bigg[it_2Q_0\s\sum_{k=0}^{K+1}(-r_1r_2Q_0^2\ss)^k&\cr
&\times\sum_{n=1}^N\{{}^{n+k-1}C_k\}(r_1Q_1\c
)^{n-1}\sum_{m=1}^M\{{}^{m+k-1}C_k\}(r_2Q_2\c)^{m-1}\bigg]&\cr
&+a_0\bigg[t_1Q_1\c\sum_{p=1}^j(r_1Q_1\c)^{p-1}
-t_1r_2Q^2_0\ss\sum_{k=0}^K(-r_1r_2Q_0^2\ss)^k&\cr
&\times\sum_{n=1}^N\{{}^{n+k}C_{k+1}\}(r_1Q_1\c
)^{n-1}\sum_{m=1}^M\{{}^{m+k-1}C_k\}(r_2Q_2\c)^{m-1}\bigg],&(3a)\cr
y_4[j]=& b_0\bigg[t_2Q_2\c\sum_{p=1}^j(r_2Q_2\c)^{p-1}
-t_2r_1Q_0^2\ss\sum_{k=0}^K(-r_1r_2Q_0^2\ss)^k&\cr
&\times\sum_{n=1}^N\{{}^{n+k-1}C_{k}\}(r_1Q_1\c
)^{n-1}\sum_{m=1}^M\{{}^{m+k}C_{k+1}\}(r_2Q_2\c)^{m-1}\bigg]&\cr
&\quad +a_0 \bigg[it_1Q_1\s\sum_{k=0}^{K+1}(-r_1r_2Q_0^2\ss)^k&\cr
&\times\sum_{n=1}^N\{{}^{n+k-1}C_k\}(r_1Q_1\c
)^{n-1}\sum_{m=1}^M\{{}^{m+k-1}C_k\}(r_2Q_2\c)^{m-1}\bigg],&(3b)\cr}$$
where
$$\eqalignno{
Q_0=&\exp[ik_0(2L_a+l_1+l_2)]&\cr
Q_1=&\exp[ik_0(2L_a+2l_1)]&(3c)\cr
Q_2=&\exp[ik_0(2L_a+2l_2)]&\cr}$$
are the phase factors which satisfy $Q_0^2=Q_1Q_2$. The upper limits of
the integers $k$, $n$ and $m$ are represented by $K$, $N$, and $M$
respectively. For any value of $j$, we obtain all the terms in the series
that are characterised by sets of values of these limits satisfying the
following relationship:
$$j=2K+M+N.\eqno(3d)$$
So, in the complete expression for $x_4$, say, there would be several
terms each involving the product of three summations over $k$, $m$ and
$n$, one term for each combination of $K$, $M$ and $N$ satisfying Eq.(3d).
So, as $j$ increases, the number of triple products of summations in
Eqs.(3a) and (3b) also increase.
The explanation of how these upper limits arise is provided below.

In this connection, interested readers may also like to know the physical
interpretation of all the terms appearing in series (3a) and (3b). One may
note that any light gets multiplied by either ($i\s$) or ($\c$) depending
on whether it changes its cavity (from PRC to SRC and {\it vice versa}) or
not after passing through the BS and one arm.

For example, we may consider the series in the second term of the
coefficient of $a_{0}$ in the intra-cavity field $x_4[j]$. Any term in
this series represent the contribution of $a_0$ light in $x_4$ after it
completed $(k+n)$ number of round trips in PRC and ($k+m$) number of round
trips in SRC. Since $j$ represents the sum of the maximum number of round
trips in two cavities, this explains why we obtain the relation (3d) for
the series described in Eqs.(3a) and (3b).
The number of possible ways of completing such a trip would be
$\{{}^{n+k}C_{k+1}\}\{{}^{m+k-1}C_k\}.$
As an easy example, let us take the contribution corresponding to
$k=0$, $n=2$, $m=1$, i.e.,
$-2r_1r_2t_1\c\ss,$
which may be expanded as
$$(r_1\c)(i\s)r_2(i\s)t_1+(i\s)r_2(i\s)(r_1\c)t_1.$$
This means that this term has been contributed by two parts of $a_0$ light
which made two round trips of PRC and one of SRC in the following two ways:
\item{(i)} \r PRM; transmitted inside \r BS, arm \r SRC \r SRM; reflected
\r BS, arms \r PRC \r PRM; reflected \r BS, arm \r PRC.
\item{(ii)} \r PRM; transmitted inside \r BS, arm \r PRC \r PRM; reflected
\r BS, arms \r SRC \r SRM; reflected \r BS, arm \r PRC.

The intra-cavity fields, $x_4$ and $y_4$ achieve their steady state value
as $j$  tends to infinity. Now, identifying
$$\sum_{n=1}^\infty\{
{}^{n+k-1}C_k\}(r_1Q_1\c)^{n-1}={1\over(1-r_1Q_1\c)^{k+1}},\eqno(4)$$
etc. and after doing some algebraic manipulations, we finally arrive at
the steady state expressions:
$$\eqalignno{
x_4[{\rm steady}]&=x_4[j\to\infty]={1\over
\chi}[a_0t_1Q_1(\c-r_2Q_2)+ib_0t_2Q_0\s],&(5a)\cr
y_4[{\rm steady}]&=y_4[j\to\infty]={1\over
\chi}[ia_0t_1Q_0\s+b_0t_2Q_2(\c-r_1Q_1)],&(5b)\cr
}$$
where
$$\chi=1-(r_1Q_1+r_2Q_2)\c+r_1r_2Q_1Q_2.\eqno(5c)$$
The steady-state expressions (5) can, of course, be calculated very
easily by assuming that, under steady-state condition, the change in
the intra-cavity fields after every
round-trip of the cavities is negligibly small [7]. One can then write
the following two equations which can be solved to obtain the
above-mentioned expressions.
$$\eqalignno{x_4[{\rm steady}]=& a_0t_1Q_1\c+x_4[{\rm steady}]r_1Q_1\c&\cr
&+ib_0t_2Q_0\s+iy_4[{\rm steady}]r_2Q_0\s,&(6a)\cr
y_4[{\rm steady}]=&b_0t_2Q_2\c+y_4[{\rm steady}]r_2Q_2\c&\cr
&+ia_0t_1Q_0\s+ix_4[{\rm steady}]r_1Q_0\s.&(6b)\cr}$$
The intra-cavity fields can achieve their maximum values at the resonance
condition of the dual recycling cavity, i.e., when $l_1=l_2$ and
$Q_0=Q_1=Q_2=1$. This implies that the intra-cavity fields $x_4$ and $y_4$
add up in phase with the incoming light. For a sufficiently low value of
$r_2$, this condition also represents {\it the broadband mode of operation}
of a dual recycled cavity [4]. The expressions for the case of power
recycling (when no SRM is used) can be easily obtained by setting $r_2=0$.
Similar exercise can of course be carried out for interferometers
incorporating Fabry-Perot cavities in the arms.

For exactly equal armlengths of the interferometer ($\theta=0$), the dual
recycling cavity, at any frequency, gets decoupled into two equivalent
three mirror cavities. If $Q_1=1$, the laser light, $a_0$ becomes resonant
with the three mirror power recycling cavity (PRM and two EMs), whereas
the sidebands of laser light (generated by the gravitational wave[3])
travel down the SRC. One can make one of these sidebands (say,
$\omega_0+\omega_g$) resonant with the three mirror signal recycling
cavity (SRM and two EMs) by adjusting $l_2$, such that
$Q_2\exp[+i\omega_g(2L_a+2l_2)]=1$. This is called {\it the narrowband mode
of operation}. Obviously, under this mode of operation, neither the $b_0$
light nor its sidebands ( which, after getting created, travel down the PRC)
can become resonant with any of the cavities.

Gravitational radiation from a coalescing compact binary system has a
{\it chirp} waveform which continually increases in frequency and amplitude
as the two stars spiral in towards each other. To observe this broadband
waveform, dynamical dual recycling technique[8] was proposed, which
incorporated two complementary techniques to improve the sensitivity of
observation: (i) narrow band observation to achieve high signal to noise
ratio at a given frequency, and (ii) tracking the chirp frequency by
adjusting the tuning frequency by changing $l_2$ as long as the signal
lies within the detector bandwidth. To implement this technique, it is
important to understand how exactly the signal (i.e., the sideband that
is tracked down) builds up its power in the presence of errors in
tuning the SRC. A programme simulating this technique in the way described
above, therefore, may come to be handy and useful.

 In actual LIGO type interferometers, the Fabry Perot cavities in the
arms may be physically different in length by several meters, and the
storage times in the arms different by as much as 1.0\% due to differing
mirror transmissions or deviations from the operating point [9]. It may
be noted that this way of doing the calculation exactly tracks down
various light beams and thus provides an answer to the question  `how many
times a particular beam traverses through any particular arm or cavity
before a specific point of time in a transient state'. The arbitrary
phase noise (mainly caused by various nonlinear effects, e.g. scattering,
thermoelastic effects on the mirrors etc.) induced on the beams in
different arms may lead to leakage of laser light to the SRC.
Also, an important  issue is the control of the resonance conditions
in the various cavities in these type of interferometers made of
suspended mirrors. For this purpose it is useful to compute the
transient behaviour of the light amplitudes at various locations in the
interferometer for given motions of mirrors. This particular algorithm
will help in investigating such kinds of transient state of the
interferometer and in tracking down the contribution of different
cavities in the total phase noise in an effective way. Investigations
are currently in progress in this line and results will be communicated
in future.

One point of caution may be worth noting here, although that is not
important for the present designs of the interferometric gravitational
wave detectors. If the Fabry Perot cavities in the arms of the
interferometer are significantly different in length, then within a given
time period, the number of bounces of light in one cavity will become
different from the number of bounces in the other. The analysis in the
form presented here will break down in that case because the iteration
parameter $j$ will differ for different beams depending on the path
history.

We point out again that the series compaction to arrive at Eqs.(3a) and
(3b) was performed by {\it manual} inspection of the recursive series
expansion produced by {\it Mathematica}. Employing {\it Mathematica} to
find these automatically is, in general, impossible since there may be
infinite number of functions with identical series expansions up to a
certain degree. Abbott[10] has studied different aspects of this
{\it Mathematica} problem and  showed a method of compacting  series
using generating functions, if the corresponding  recurrence relation
is available. Investigation is in order to find a way to combine the
complicated recursive process of the present problem with the method
described by Abott[10] for simpler cases, with the purpose of employing
{\it Mathematica} to arrive at Eqs(3a) and (3b) automatically. It should
be noted, however, that this is not a problem for the applicability of
the method of calculation described in this paper to realistic
situations where one has to do numerical computation.

\vskip.2truein

It is a pleasure to thank S.V. Dhurandhar and B.S. Sathyaprakash for
discussions and Paul Abbott for taking active interest in the problem
of `series compaction' on my request.
\vfill\eject
\n{\bf References}

\i{1.} A. Abramovici, W.E. Althouse, R.W.P. Drever, Y. G\"ursel, S.
       Kawamura, F.J. Raab, D. Shoemaker, L. Sievers, R.W. Spero,
       K.S. Thorne,
       R.E. Vogt, R. Weiss, S.E. Whitcomb, and M.E. Zucker,
       ``LIGO : The Laser Interferometer Gravitation-wave Observatory",
       Science, {\bf 256}, 325-333 (1992)
\i{2.} R.W.P. Drever, ``Interferometer detectors for gravitational radiation"
       in {\it Gravitational Radiation}, N. Deruelle and T. Piran,
       eds. (North-Holland, Amsterdam, 1983), pp. 321-338.
\i{3.} J.-Y. Vinet, B.J. Meers, C.N. Man, A. Brillet,
       ``Optimization of long-baseline optical interferometers for
       gravitational-wave detection,"
       Phys. Rev., {\bf D38}, 433-447 (1988).
\i{4.} B.J. Meers, ``Recycling in laser-interferometric
       gravitational-wave  detectors,"
       Phys. Rev., {\bf D38}, 2317-2326 (1988);
       {\it ibid}, ``The frequency response of interferometric gravitational
       wave detectors,"
       Phys. Lett., {\bf A 142}, 465-470 (1989).
\i{5.} S. Wolfram, {\it Mathematica: a system for doing mathematics by
       computer} (Addison-Wesley Publishing Co., California, 2nd ed., 1991)
\i{6.} C.M. Caves, ``Quantum-mechanical noise in an interferometer,"
       Phys. Rev., {\bf D23}, 1693-1708 (1991).

\i{7.} A. Brillet, J. Gea-Banacloche, G. Leuches, C.N. Man, J.-Y. Vinet,
       ``Advanced techniques : recycling and squeezing," in
       {\it The detection of gravitational waves}, D.G. Blair, ed.
       (Cambridge University Press, Cambridge, 1991), pp. 369-405;
       V. Chickarmane and B. Bhawal, ``Squeezing and dual recycling in laser
       interferometric gravitational wave detectors,"
       Phys. Lett., {\bf A190}, 22-28 (1994).
\i{8.} B.J. Meers, A. Krolak, J.A. Lobo, ``Dynamically tuned interferometers
       for the observation of gravitational waves from coalescing compact
       binaries,"
       Phys. Rev., {\bf D47}, 2184-2197 (1993).
\i{9.} D. Shoemaker, LIGO project, Massachusetts Institute of Technology,
       Cambridge, MA02139, USA (personal communication, 1995).
\i{10.} P. Abbott, ``Automatic summation," The Mathematica Journal, vol.5,
       issue 3, 33-35 (1995).

\vfill\eject
\n{\bf Figure Caption}
\vskip.2truein
\item{FIG.1}: Optical arrangement for dual recycling. (BS) - Beam Splitter,
(EM) - End Mirror, (PRM) - Power Recycling Mirror, (SRM) - Signal Recycling
Mirror, (PD) - Photodetector. The light beams: $a_0$ - input laser beam,
$b_0$ - ordinary or squeezed vacuum state ({\it see text}); $x_1$, $x_4$ -
intracavity fields in power recycling cavity; $y_1$, $y_4$ - intracavity
fields in signal recycling cavity; $x_1$, $x_2$, $y_2$, $y_3$ - light
fields inside the arms;     $a_1$, $b_1$ - output beams through PRM and
SRM respectively.
\item{FIG.2}: Mirror and light beams. $a_0$, $b_0$ - input beams; $a_1$,
$b_1$ - output beams.
\item{FIG.3}: Flowchart of a Fabry-Perot cavity at non-steady state. CM -
Corner Mirror, EM - End Mirror. The circle with $L$ inside represents
phase change due to traversal of cavity length, $L$.

\item{FIG.4}: Flowchart of a single delay-line dual recycled cavity
at non-steady state. PRM - Power Recycling Cavity, SRM - Signal
Recycling Cavity, BS - Beam Splitter.  Circle with $l_1$ or $l_2$ inside
represents phase changes due to traversal of length, $l_1$ or $l_2$ between
BS and PRM or SRM respectively.

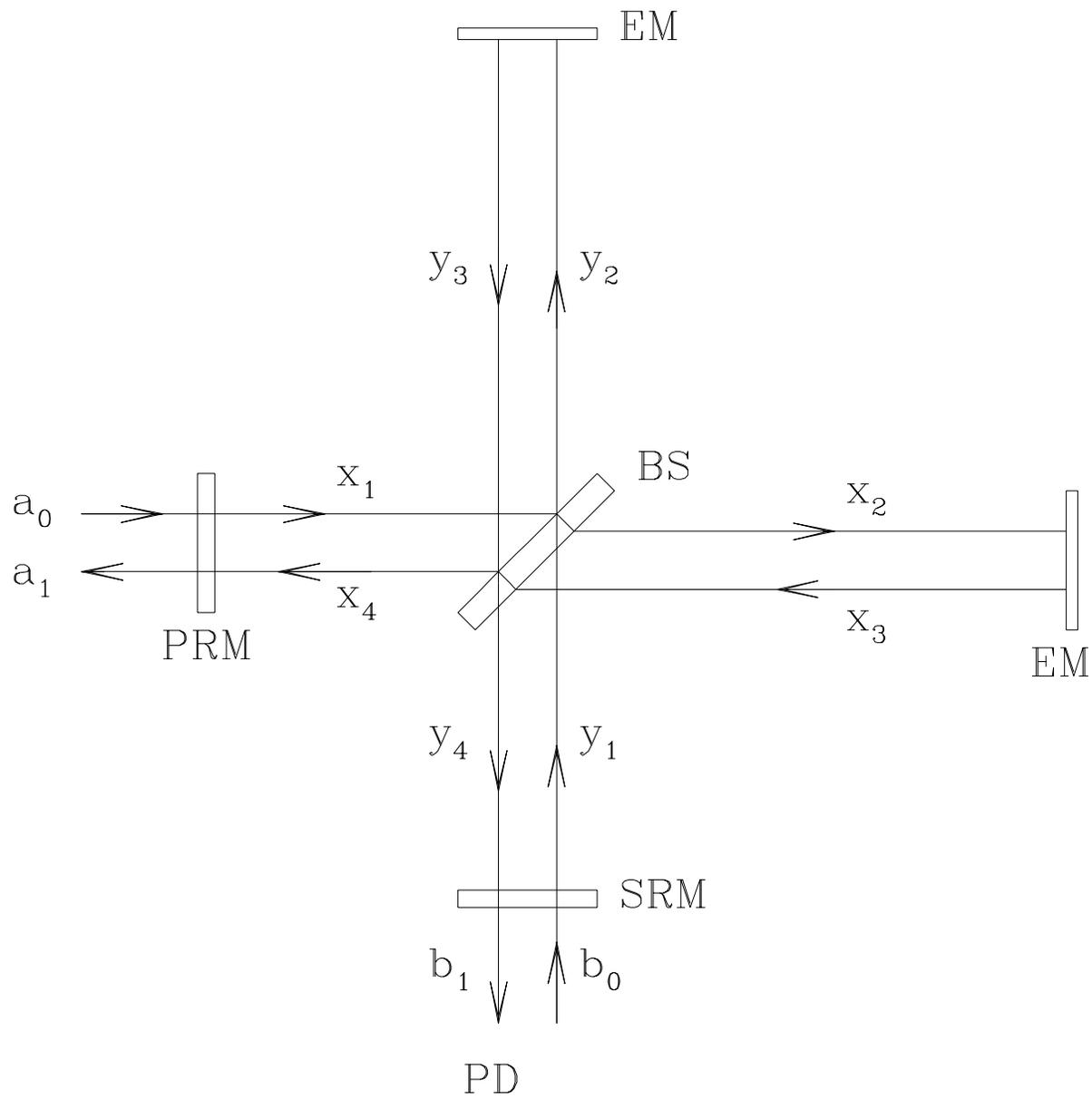
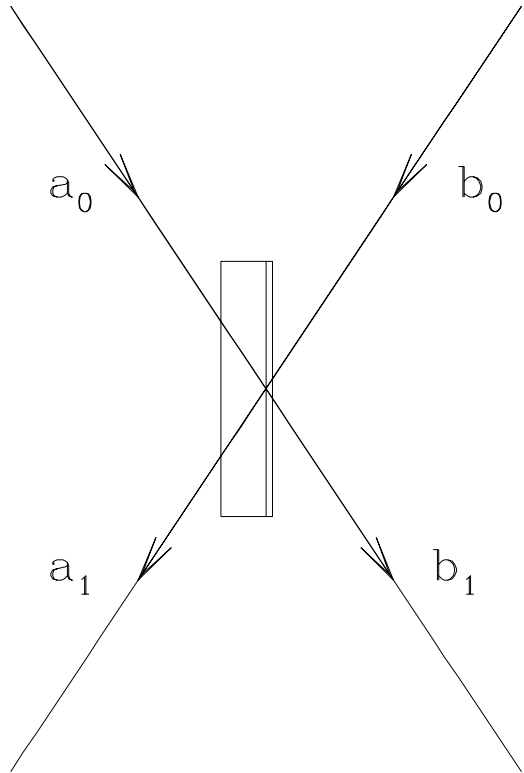
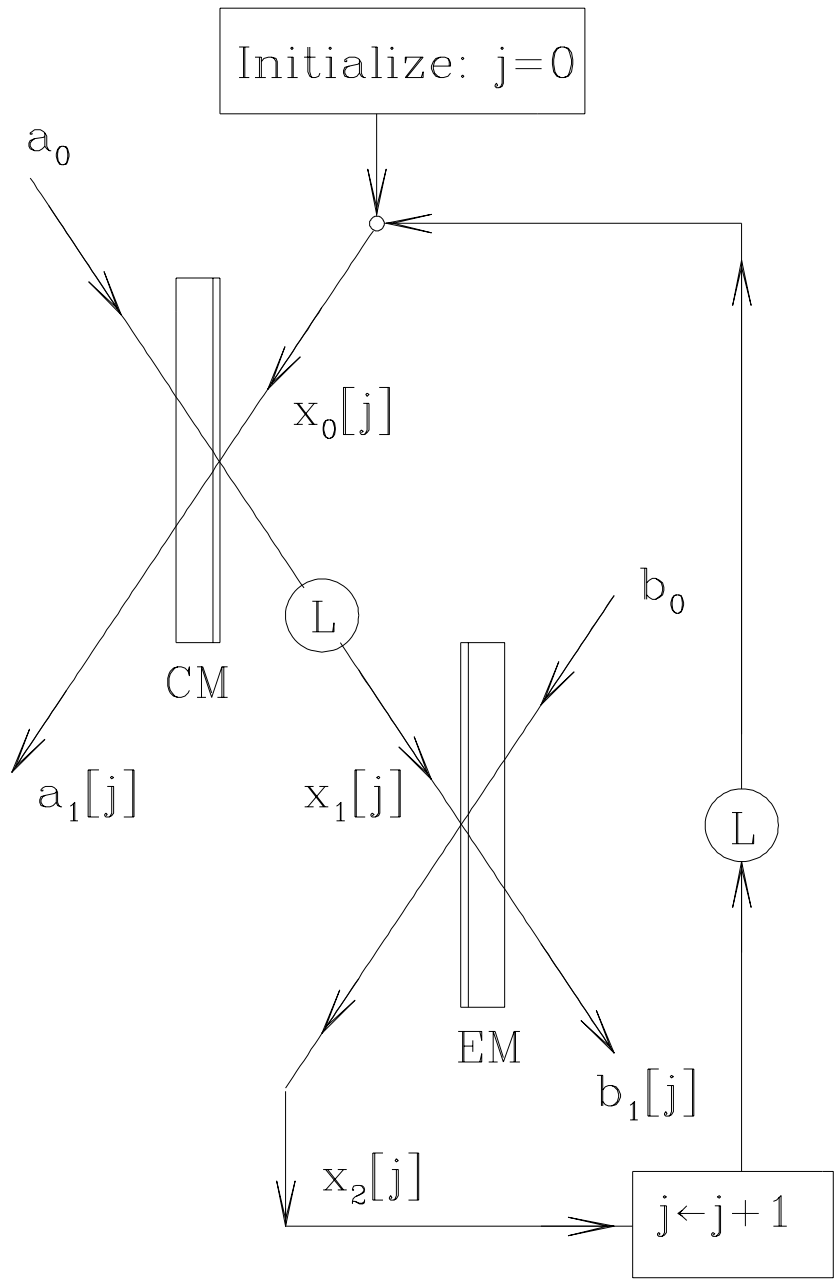
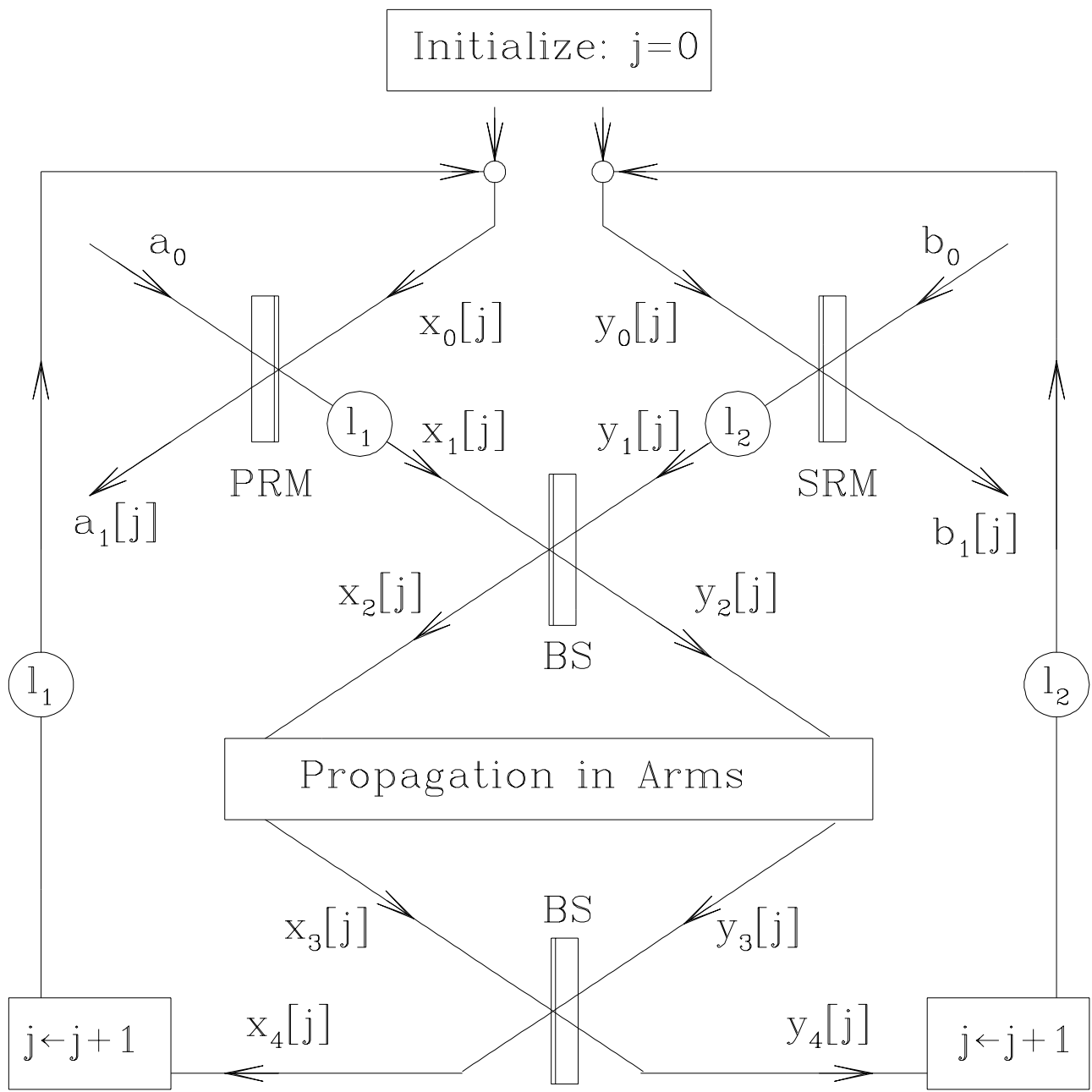
\end